 \documentclass[final,5p,times,twocolumn]{elsarticle}

 \usepackage{graphics,graphicx}
\usepackage{amssymb}
\journal{Physics Letters A}

\begin{document}

\begin{frontmatter}



\title{Excitability in optical systems close to $Z_2$-symmetry}


\author[label1,label2]{Stefano Beri \corref{cor1}\fnref{footnote}}
\author[label1,label2]{Lilia Mashall}
\author[label1]{Lendert Gelens}
\author[label1]{Guy Van~der~Sande}
\author[label3]{Gabor Mezosi}
\author[label3]{Marc Sorel}
\author[label1,label2]{Jan Danckaert}
\author[label1]{Guy Verschaffelt}
\address[label1]{Department of Applied Physics and Photonics, Vrije Universiteit Brussel, Pleinlaan 2, 1050 Brussels, Belgium}
\address[label2]{Department of Physics, Vrije Universiteit Brussel, Pleinlaan 2, 1050 Brussels, Belgium}
\address[label3]{Department of Electronics \& Electrical Engineering, University of Glasgow, Rankine Building, Oakfield Avenue,
Glasgow, G12 8LT, United Kingdom}
\fntext[footnote]{Author to whom correspondence should be addressed; stefano.beri@vub.ac.be}
\begin{abstract}
We report theoretically and experimentally on excitability in semiconductor ring lasers 
in order to reveal a mechanism of excitability, general for systems close to $Z_2$-symmetry.
The global shapes of the invariant manifolds of a saddle in the vicinity of a homoclinic loop determine the origin of excitability and the features of the excitable pulses.
We show how to experimentally make a semiconductor ring laser excitable by breaking the $Z_2$-symmetry in a controlled way. The experiments confirm the theoretical predictions.
\end{abstract}

\begin{keyword}
Semiconductor Ring Laser \sep Excitability \sep Fluctuations \sep Nonlinear Dynamics \sep Stochastic Dynamics
\PACS 05.40.-a,42.55.Px,42.60.Mi

\end{keyword}

\end{frontmatter}

% \linenumbers

%%\section{} \label{}
Today, excitability in systems outside equilibrium is a very fertile and interdisciplinary research topic.
Excitability has been observed in various fields including Chemistry, Physics, Biology or neural networks for computation tasks \cite{Lindner04a,Pikovsky97a,Kapral95a,Bertram95a,Hodgkin52a,Izhikevich00a,Maass97a}.
In particular, excitability attracts a lot of attention in the field of optics \cite{Yacomotti94a,Dubbeldam97a,Wieczorek02a,Gomila05a,Marino05a,Wunsche02a,Giacomelli00a,Goulding07a} due to its application as a way to generate well defined optical pulses and therefore starting a quest for optical excitable units.
Lasers with saturable absorber \cite{Yacomotti94a,Dubbeldam97a}, optically injected lasers \cite{Goulding07a,Wieczorek02a}, lasers with optical feedback \cite{Giacomelli00a} or VCSELs with opto-electronic feedback \cite{Romariz07a} have all been proposed as optical excitable units, witnessing both the high interest on optical excitability as well as the difficulty in achieving an ultimate design.

The aim of this paper is two-fold: we introduce a simple, integrable and scalable optical excitable unit based on a semiconductor ring laser (SRL) and we disclose in general the excitable properties of the wide class of nonlinear dynamical systems with weakly broken $Z_2$-symmetry \cite{Wiesenfeld82a,Kuznetsov04a}.
SRLs are semiconductor lasers whose cavity has a circular geometry that can sustain operation in two counterpropagating directions, namely clockwise (CW) and counter-clockwise (CCW) \cite{Krauss95a,Sorel03a}.
It was shown both theoretically \cite{VanderSandeJPhysB2008} and confirmed experimentally \cite{Gelens2009b, Beri08a} that the dynamical regimes of a symmetric SRL are described by a two-dimensional asymptotic system of the form: 
\begin{equation}
\dot{\theta} = f_1\left( \theta , \psi \right) \; \; \; \; 
\dot{\psi}   = f_2\left( \theta , \psi \right), \label{Eq:Z2}
\end{equation}
with the vector field $f_{1,2}\left( \theta , \psi \right)$ being invariant under the transformation $\theta \to -\theta$, $\psi \to 2\pi-\psi$, making SRLs an optical prototype of $Z_2$-symmetric systems.

For the majority of optical systems, excitability takes place around a homoclinic bifurcation of a {\it stable} limit cycle \cite{Goulding07a,Wieczorek02a,Yacomotti94a,Dubbeldam97a,Gomila05a}. The presence of the stable limit cycle can easily be observed in these systems as it leads to Q-Switching oscillations. However, it is known that only {\it unstable} limit cycles undergo homoclinic bifurcation in SRLs \cite{Gelens09a}, which are intrinsically more challenging to address as they are not associated to observable dynamical regimes. 

In the first part of this paper, we present the theory for excitability in SRLs. Our approach is based on the investigation of the global topology of the invariant manifolds in 
systems with a slightly broken $Z_2$-symmetry.
In such a way we can predict the onset of excitability near the homoclinic bifurcation of an {\it unstable} limit cycle as well as the properties of single and multiple excitable pulses.
In the second part of this paper we show how to experimentally break the $Z_2$-symmetry in a SRL in a controlled way. Excitability is revealed, whose properties confirm the topological predictions.

Consider a SRL operating in single-transverse and single longitudinal mode. Two directional modes, $CW$ and $CCW$, can operate in the ring cavity with different intensities $P_{CW/CCW}$ and phases $\phi_{CW/CCW}$. A linear coupling parameter with amplitude $K$ and phase $\phi_K$ is used to describe the transfer of power between $CW$ and $CCW$. The two directional modes operate in antiphase, conserving the total power $P_{CW}+P_{CCW}$ \cite{Sorel03a, VanderSandeJPhysB2008, Gelens2009b,Beri08a}, such that an asymptotic two-dimensional set of equations can model the SRL operation:
\begin{eqnarray}
\dot{\theta} &=& J \sin\theta\cos\theta + 2\left( 1 - \delta \right) \cos \left( \phi_k + \psi \right) \nonumber \\
&&-\left( 1 - \sin \theta \right) \left[ \left( 1 - \delta \right) \cos\left( \phi_k + \psi \right) \right. \label{eq:thetadot} \\
&& \left. +  \left( 1 + \delta \right) \cos\left( \phi_k - \psi \right)  \right] \nonumber \\
\cos\theta \dot{\psi} &=& \alpha J \sin\theta\cos\theta \nonumber \\
&&- \left( 1 + \delta \right) \left( 1 - \sin \theta \right) \sin \left( \phi_k - \psi \right) \label{eq:psidot} \\
&& + \left( 1 - \delta \right) \left( 1 + \sin \theta \right) \sin \left( \phi_k + \psi \right) \nonumber
\end{eqnarray}

Here the angular variable $\theta = 2 \arctan \sqrt{P_{CCW} / P_{CW}} - \pi/2$ quantifies the power partitioning between the counterpropagating modes. The phase difference $\psi = \phi_{CCW}-\phi_{CW}$ is the second dynamical variable in the system. $J$ is the rescaled bias current and $\alpha$ is the linewidth-enhancement factor.
The $Z_2$-symmetry of Eqs.~(\ref{eq:thetadot})-(\ref{eq:psidot}) is continuously broken by the introduction of an asymmetry $\Delta K $ in the mode-coupling. The dimensionless parameter $\delta = \Delta K / 2K$ measures the relative magnitude of the symmetry breaking. in what follows we take $\delta = 4.5\%$ which corresponds to breaking the symmetry of the system in favor of the $CW$ mode. In order to have a quantitative comparison with the experimental results shown below, we choose $K=0.1715$ $ns^{-1}$ as the dimensionless time in Eqs.~(\ref{eq:thetadot})-(\ref{eq:psidot}) is scaled with the physical backscattering strength $K$. We now discuss the phase space topology when varying the bias current $J$.

\begin{figure}[t]
\centering
\includegraphics[width=\columnwidth]{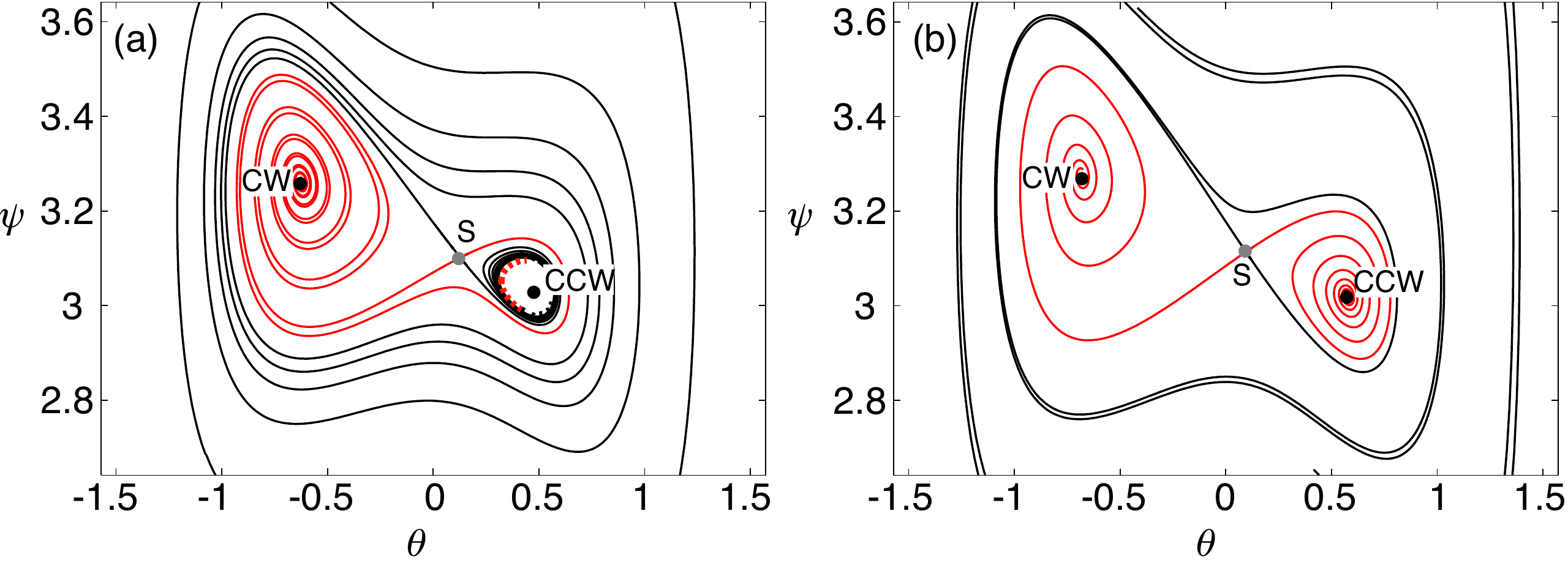}
\caption{\label{Fig:hombif} Phase space portraits of Eqs.~(\ref{eq:thetadot})-(\ref{eq:psidot}) for different values of the bias current $J$. Black lines depict the stable manifold of $S$, while the red (grey) lines represent the unstable manifold of the $S$. In (a) the unstable limit cycle is shown in the red (grey) dashed line. The parameters are the following:
$\delta = 4.5 \%$,  $\phi_K = 1.5$. (a) $J=0.659$, (b) $J=0.691$.}
\end{figure}

The phase space of the SRL consists of two stable states $CW$ and $CCW$ whose basins of attraction are separated by the two branches of the stable manifold of a saddle $S$ \cite{Beri08a,Gelens09a}.
The main changes in the topology take place 
when the current $J$ crosses a critical value $J_{hom}$ which corresponds to 
a homoclinic bifurcation of an unstable limit cycle as shown in Fig.~\ref{Fig:hombif}(a)-(b).
For $J<J_{hom}$ [Fig.~\ref{Fig:hombif}(a)], a small 
unstable limit cycle surrounds $CCW$ and prevents the operation of the SRL in the $CCW$ mode.
Both branches of the unstable manifold of $S$ connect $S$ with $CW$, whereas one branch of the stable manifold of $S$ connects $S$ with the limit cycle and the other branch with an unstable node in $\left( 0,0 \right)$ [not shown].
Excitability is possible in this scenario when the system -initially residing in $CW$- is driven by a fluctuation beyond the stable manifold of the saddle point $S$ and subsequently relaxes back following the unstable manifold of   $S$ \cite{Wunsche02a}.
The radius of the unstable limit cycle rapidly grows with the current until it disappears in a homoclinic loop at $J = J_{hom}$.
For $J>J_{hom}$, the basins of attraction of the $CW$ and $CCW$ modes are separated by the two branches of the stable manifold of the saddle, a scenario that suggests bistability [Fig.~\ref{Fig:hombif}(b)]. One would expect the system to leave the basin of attractor of the $CCW$ state and relax to the opposite mode following a branch of the unstable manifold of $S$.
\begin{figure}[t]
\centering
\includegraphics[width=\columnwidth]{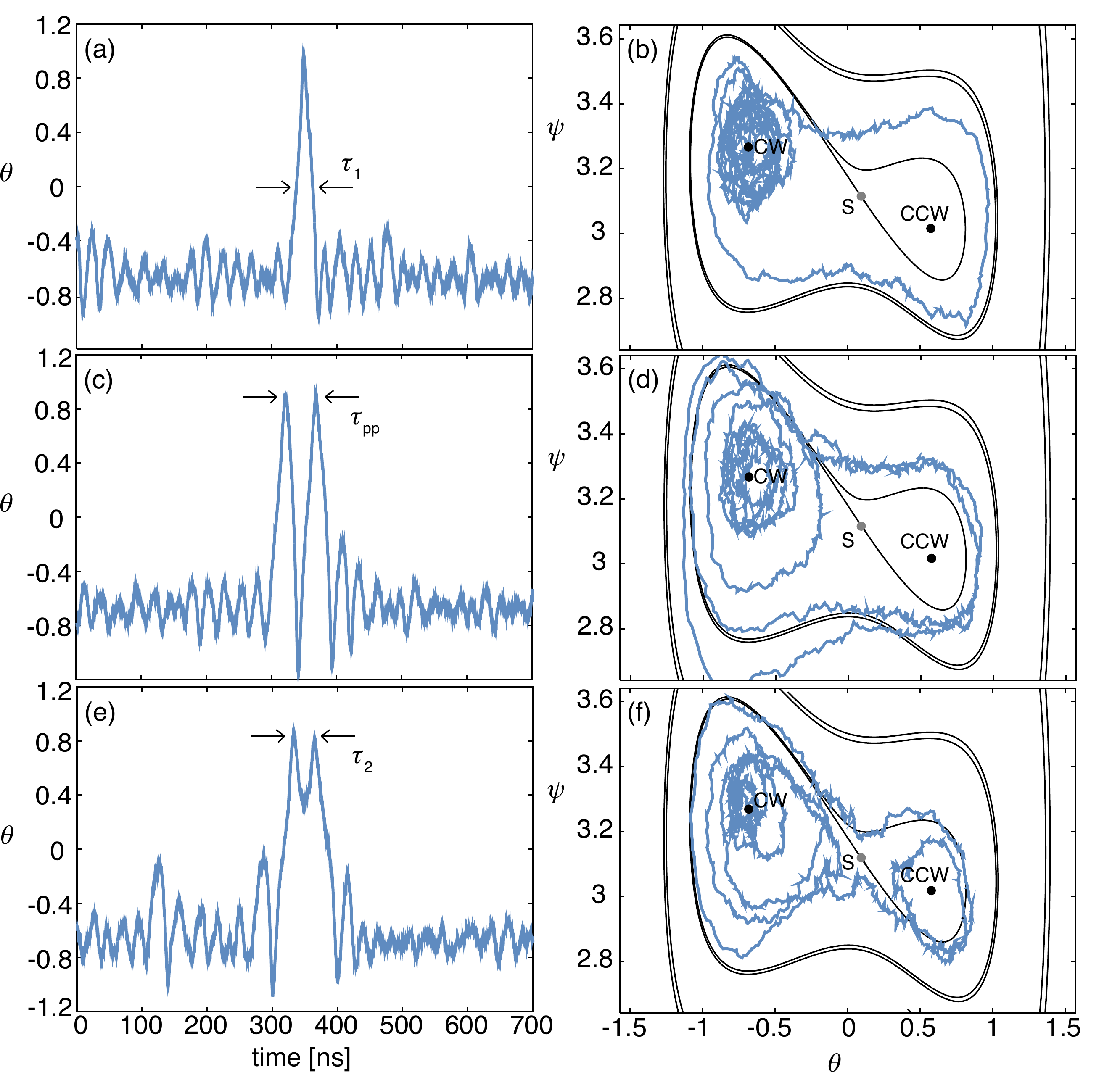}
\caption{\label{Fig:PhaseSpacePics} Numerical solutions of Eqs.~(\ref{eq:thetadot})-(\ref{eq:psidot}) revealing excitable pulses in time domain (a,c,e) and their projection on the phase space (b,d,f). (a) Single pulse; $\tau_1$: pulse width. (c) Double pulse; $\tau_{pp}$: intra-peak interval. (e) pulse with two maxima associated with the "metastability" of the $CCW$ state; $\tau_2$: distance between the two maxima. The parameters are the same as in Fig.\ \ref{Fig:hombif}(b).}
\end{figure}
However, close to the homoclinic bifurcation, the global shape of the invariant manifolds is reminiscent of the homoclinic loop.
As it is shown in Fig.~\ref{Fig:hombif}(b), the unstable manifold of $S$ then spirals very slowly to the $CCW$ mode. The distance between consecutive laps of the unstable manifold decreases when $J \to J_{hom}$. 
Close to the homoclinic loop,
the system relaxes very slowly to the $CCW$ state, therefore making the $CCW$ operation metastable: noise fluctuations will remove the system from the basin of attraction of $CCW$ before $CCW$ operation can be established.

In the same way, close to the homoclinic bifurcation, the distance between the branches of the stable manifold of the saddle becomes negligible 
when compared to the diffusion length-scale induced by the noise. 
The generation of an excitable pulse therefore corresponds to a noise activated crossing of two arbitrarily close thresholds.
An example of an excitable pulse in time domain is given in Fig.~\ref{Fig:PhaseSpacePics}(a) and the corresponding phase space trajectory is shown in Fig.~\ref{Fig:PhaseSpacePics}(b).
Opposite to the scenario in Fig.~\ref{Fig:hombif}(a), here the unstable manifold of $S$ plays no role in the excitability. Excitability disappears for $J$ too small or too large. When $J$ is too large, the metastability of the $CCW$ turns into full stability such that the system becomes bistable. For $J$ too small, the system will exhibit alternating oscillations \cite{Gelens09a}.

For both $J<J_{hom}$ and $J>J_{hom}$, we expect the intra-spike-interval (ISI) between consecutive excitable events to be distributed in an exponential Arrhenius way as for a standard activation problem \cite{Melnikov91a}.
The duration $\tau_1$ of the pulse is instead a deterministic time scale corresponding to a revolution of the system in the phase space [see Fig.~\ref{Fig:PhaseSpacePics}(b)]. 
The peak-width has been calculated from numerical integration of Eqs.~(\ref{eq:thetadot})-(\ref{eq:psidot}) by averaging the Full-Width-Half-Maximum of a sample of 771 peaks yielding $\tau_1 = 27.54$ns.

The shape and the features of the excitable pulses close to $Z_2$-symmetry can also be explained by the topology of the invariant manifolds.
Consider the double pulses shown in  Fig.~\ref{Fig:PhaseSpacePics}(c): two well defined pulses are emitted by the SRL with an average peak-to-peak interval $\tau_{pp}=49.22$ns (averaged on a sample of 286 double pulses).
Double pulses can be explained as follows. 
The emission of the first pulse is due to a noise induced activation;
during the pulse, the system evolves deterministically towards the $CW$ state. 
When the $Z_2$-symmetry is weakly broken and $J$ is close to $J_{hom}$, 
the deterministic evolution of the pulse brings the system in the vicinity of the stable manifold of $S$. A second excursion in the phase space, without residence in the $CW$ mode is therefore possible as shown in Fig.~\ref{Fig:PhaseSpacePics}(d).
The peak-to-peak interval for such double pulses is determined by the duration of the deterministic rotation around the saddle in the phase space which is approximately twice the pulse-width.
This topological insight is confirmed by the almost $2:1$ relation between $\tau_{pp}=49.22$ns and the pulse-width $\tau_1=27.54$ns.
We remark that deterministic mechanisms for the generation of multiple pulses such as those in \cite{Goulding07a,Wieczorek02a} require bifurcation scenarios that cannot exist in two-dimensional systems and therefore cannot be observed for Eqs.~(\ref{eq:thetadot})-(\ref{eq:psidot}) and in general for $Z_2$-planar systems.

Finally, excitable events occur that are characterized by two maxima separated by a minimum that does not reach the ground state such as the one shown in Fig.~\ref{Fig:PhaseSpacePics}(e).
The presence of such pulses is evidence of the metastability of the $CCW$-mode as discussed above:
if the system enters the basin of attraction of $CCW$, it starts a rotation around $CCW$ which results in the oscillation of power. Due to the closeness to the homoclinic bifurcation, during the rotation the system remains close to the stable manifold of $S$. 
The noise eventually drives the SRL back into the basin of attraction of the $CW$ mode
as shown in Fig.~\ref{Fig:PhaseSpacePics}(f).
The interval between the two peaks corresponds to a rotation around the $CCW$ node and can be estimate as being $1/2$ of the period of a full rotation around the phase space.

\begin{figure}[t!]
\centering
\includegraphics[width=\columnwidth]{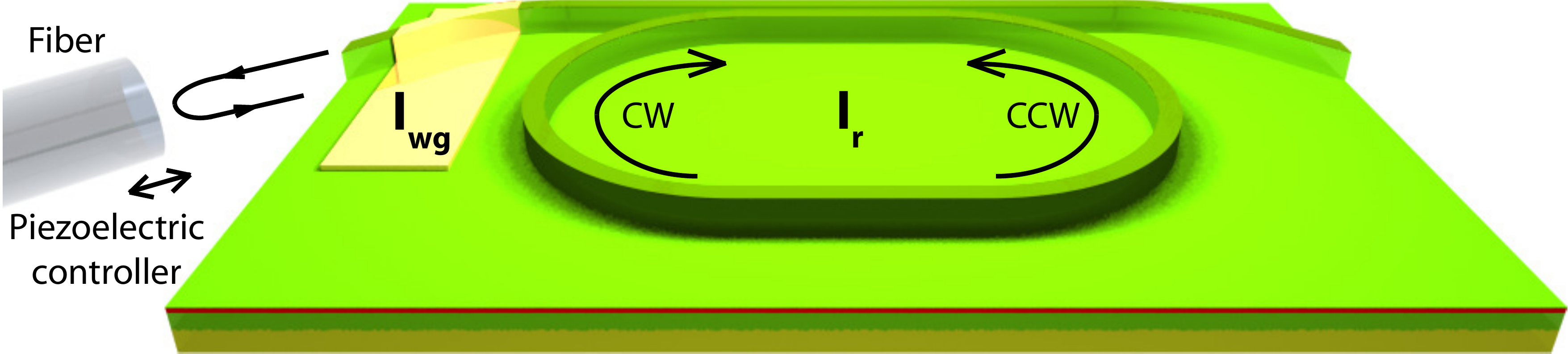}
\caption{Schematic of the SRL and the experimental setup. The notation is as follows. $CW$/$CCW$: counterpropagating modes; $I_{wg}$ bias current on the waveguide; $I_r$: bias current on the ring.
 \label{figurechip}}
\end{figure}

The experiments have been performed on an InP-based multiquantum-well SRL with a racetrack geometry and a free-spectral-range of $53.6$ GHz. The device operates in a single-transverse, single-longitudinal mode at $\lambda = 1.56 \mu$m.
The power is extracted from the ring cavity by directly coupling the ring to a bus waveguide, and collected at the chip facets with a cleaved optical fibre as shown in Fig.~\ref{figurechip}.
An electrical contact has been applied to the bus waveguide which can be independently pumped.
The purpose of the contacting is twofold: it allows to counteract absorption and to amplify on-chip the signal emitted by the ring; more interestingly
the presence of a contact allow us to continuously break the symmetry of the device in a controlled way by changing in an asymmetric way the relative strength $\delta$ and phase $\Delta \phi_k$ of the coupling between $CW$ and $CCW$.
Using the cleaved facet of the fibre as a mirror, 
we are able to reflect power from one mode (for instance $CCW$) back into the waveguide and finally to the counterpropagating mode in the ring. 
The amount of power that is coupled to the $CW$ mode can then be controlled by tuning the current $I_{wg}$ on the waveguide, whereas its phase can be tuned by positioning the fibre facet with a piezoelectric controller.
The chip containing the SRL is mounted on a copper mount and thermally controlled by a Peltier element which is stabilized at a temperature of $21.00 ^\circ$C with an accuracy of $0.01^\circ$C.
We analyse the output power of the $CCW$ mode using a fast photodiode connected to an oscilloscope with a sampling rate of $4.0$ns. This sampling rate is slow enough for the purpose of recording long time series for statistical purposes, and it is fast enough to sample accurately the individual excitable pulses.

\begin{figure}[t!]
\centering
\includegraphics[width=\columnwidth]{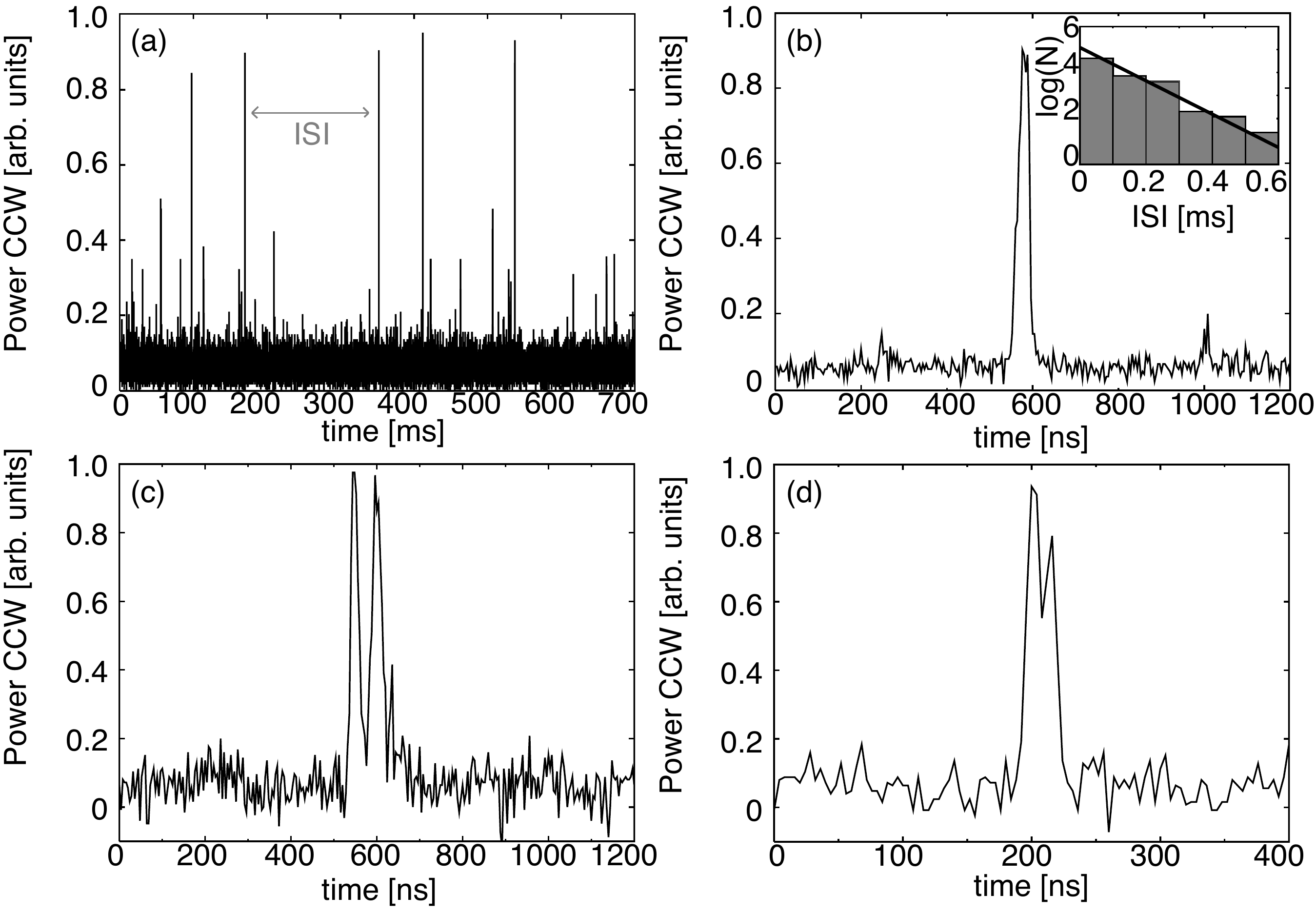}
\caption{Experimental characterization of excitability in SRLs. (a) Time trace displaying excitability for $I_r = 45.38$mA (b) detail of (a) showing a single excitable pulse; (c) double pulse excitability for $I_r = 44.62$mA; (d) metastability for $I_r = 44.62$mA. Inset of (b): distribution of the intra-spike intervals for $I_r = 45.38$mA.
 \label{Fig:experiments}
}
\end{figure}

Due to the intrinsic symmetry of SRLs, the device is expected to operate with equal probability in either the CW or the CCW state. 
However, by breaking the symmetry using the piezo-controlled optical fibre and the bias-current on the bus waveguide, the operation in one of the two states (for instance $CW$) can be favored [see Fig.~\ref{figurechip}].
Time series of the power emitted in the $CCW$ direction are shown in Fig.~\ref{Fig:experiments}(a) for a  bias-current on the ring of $I_{r} = 45.38$mA and for a waveguide current of $I_{wg} = 14.0$mA. 
It is clear from Fig.~\ref{Fig:experiments}(a) that the ring operates most of the time in the $CW$ mode.
Short excitable pulses such as the one shown in Fig.~\ref{Fig:experiments}(b) are observed.
The average width of these pulses is 
$33.9 \pm 4$ns, which agrees with the simulated value $\tau_1=27.54$ns.
The ISI are distributed exponentially (see inset in Fig.~\ref{Fig:experiments}(b)) with an average value of $154 \mu$s.
We can therefore deduce that the pulses have the characteristic signatures of excitability, being generated in noise-activation process across a threshold and having thereafter a deterministic evolution.
We remark here that no residence in the $CCW$ state was observed.

The dependence of the $ISI$ and pulse width $\tau_1$ on $I_r$ is shown in Tab.~\ref{tab:ISI}; the $ISI$ increases with $I_r$, consistent with the picture of a noise activation process, whereas the pulse width remains almost constant.

\begin{table}%[H] add [H] placement to break table across pages
\center
%\begin{ruledtabular}
\begin{tabular}{|c|c|c|c|c|}
% Lines of table here ending with \\
\hline
$I_{r}$ [mA] & ISI [$\mu$s] & $\tau_1$ [ns] & $\tau_{pp}$ [ns] & $\tau_2$ [ns] \\
\hline
44.07 & 23.3 & 30 & 53.0 & 25.0\\
44.62 & 95.20 & 26.7 & 51.0 & 26.6\\
45.07 & 110.7 & 32.5 & 49.5 & 25.5\\
45.38 & 154.6 & 33.9 & X & 28.0\\
\hline
\end{tabular}
%\end{ruledtabular}
\caption{\label{tab:ISI} Measured peak properties for different values of the bias current $I_r$ on the ring. 
The current on the waveguide for $I_p=45.07$mA is $I_{wg}=14.45$mA; for the other bias currents $I_{wg}=14.0$mA. The uncertainty is $\pm 0.01$mA for the bias current and $\pm 4$ns for the other quantities. For $I_r = 45.38$mA no double peaks were observed.}
\end{table}

Among the pulses in the time series, double-pulse events are observed such as the one shown in Fig.~\ref{Fig:experiments}(c). 
This kind of events consists of two well defined consecutive pulses separated by a drop of power to the ground state. The peak to peak interval $\tau_{pp}$ for different values of the bias current is shown in Tab.~\ref{tab:ISI}.
Other pulse-shapes such as the one in Fig.~\ref{Fig:experiments}(d)
have also been observed. They are characterized by two well-defined maxima separated by a shallow dip, and are a signature of the metastability of the $CCW$ mode as discussed above. The interval $\tau_2$ between the consecutive maxima is shown in Tab.~\ref{tab:ISI}.\\
We observe from Tab.~\ref{tab:ISI} that the ratio between the experimental values of $\tau_1$ and $\tau_{pp}$ is in agreement with the $1:2$ ratio predicted by our topological arguments. 
In the same way, the ratio between $\tau_{2}$ and $\tau_{pp}$ agrees well with the $1:2$ ratio predicted by the theory.

In conclusion, we have investigated 
excitability for generic planar systems close to $Z_2$-symmetry 
and disclosed how the shape of the invariant manifolds lead to noise-activated pulses.
An optical excitable unit based on this concept has been implemented using a multi quantum-well SRL with slightly asymmetric mode-coupling. 
Such unit can in principle be integrated on chip and does not require external optical injection or feedback from an external cavity.
We used a topological analysis 
to predict the features of three different types of excitable pulses, as well as quantitative relations between the relevant time-scales.
The predictions of the theory have been confirmed by the experiments.
We have numerically verified that the use of other non-$Z_2$ perturbations leads to similar results.

This work has been partially funded by the European Union under project IST-2005-34743 (IOLOS) and by the Research Foundation - Flanders. This work was supported by the Belgian Science Policy Office under grant No.\ IAP-VI10. G.VdS. and S.B. are Postdoctoral Fellows and L.G.  is a PhD Fellow of the FWO. S.B. acknowledges I.~A.~Khovanov for fruitful scientific discussion.



%% References


%\bibliographystyle{elsarticle-num}
%\bibliography{ref_PLA}

\begin{thebibliography}{00}


\bibitem{Lindner04a}
B.~Lindner,J.~Garcia-Ojalvo, A.~Neiman, L.~Schimansky-Geier, 
Phys.~Rep. 392, (2004) 321
\bibitem{Pikovsky97a}
A.~S.~Pikovsky, J.~Kurths, Phys. Rev. Lett. 78, (1997) 775
\bibitem{Kapral95a}
R.~Kapral, K.~Showalter, Chemical Waves and Patterns, Kluwer Academic Publishers, Dordrecht, 1995
\bibitem{Bertram95a}
R.~Bertram, M.~J.~Butte, T.~Kiemel, A.~Sherman,
B.~Math.~Biol. 57, (1995) 413-439
\bibitem{Hodgkin52a}
A.~L.~Hodgkin, A.~F.~Huxley, J. Physiol. 117 (1952) 500
\bibitem{Izhikevich00a}
E.~M.~Izhikevich, Int.~J.~Bifurcat.~Chaos 10 (2000) 1171
\bibitem{Maass97a}
W.~Maass, NeuralNetworks 10 (1997) 1659-1671 
\bibitem{Wieczorek02a}
S.~Wieczorek, B.~Krauskopf, D.~Lenstra,
Phys. Rev. Lett 88, (2002) 063901
\bibitem{Yacomotti94a}
M.~A.~Larotonda,A~Hnilo, J.~M.~Mendez, A.M.~Yacomotti
Phys. Rev. A 65 (2002) 033812
\bibitem{Dubbeldam97a}
J.~L.~A.~Dubbeldam, B.~Krauskopf, D.~Lenstra
Phys. Rev. E 60 (1999) 6580
\bibitem{Gomila05a} 
D.~Gomila, M.~A.~Matias, and P.~Colet,
Phys. Rev. Lett.  94, (2005) 063905
\bibitem{Goulding07a}
D.~Goulding, S.~P.~Hegarty, O.~Rasskazov, S.~Melnik, M.~Hartnett, G.~Greene, J.~G.~McInerney, D.~Rachinskii, G.~Huyet
Phys. Rev. Lett. 98 (2007) 153903
\bibitem{Giacomelli00a}
G.~Giacomelli, M.~Giudici, S.~Balle, J.~R.~Tredicce
Phys. Rev. Lett. 84 (2000) 3298
\bibitem{Marino05a}
F.~Marino and S.~Balle,
Phys. Rev. Lett. 94, (2005) 094101
\bibitem{Wunsche02a}
H.~J.~W\"unsche, O.~Brox O, M.~Radziunas, F.~Henneberger
Phys. Rev. Lett. 88 (2002) 023901
\bibitem{Romariz07a}
A.~R.~S.~Romariz, K.~H.~Wagner,
Appl. Optics  46, (2007) 4736--4745
\bibitem{Wiesenfeld82a}
K.~A.~Wiesenfeld, E.~Knobloch, Phys.~Rev.~A 26, (1982) 2946--2953
\bibitem{Kuznetsov04a}
Y.~Kuznetsov, Elements of Applied Bifurcation Theory, Springer, 3rd edition, New York, 2004
\bibitem{Krauss95a}
T.~F.~Krauss, R.~M.~{De La Rue}, P.~J.~R.~Laybourn, B.~Vogele, C.~R.~Stanley,
IEEE J. Sel. T. Quant. {\bf 1}, 757, (1995)
\bibitem{Sorel03a}
M.~Sorel, G.~Giuliani, A.~Scir\`e, R.~Miglierina, S.~Donati, P.~J.~R.~Laybourn
IEEE J. Quantum Electron. 39, (2003) 1187--1195
\bibitem{VanderSandeJPhysB2008} 
G.~Van~der~Sande, L.~Gelens, P.~Tassin, A.~Scir\`e, J.~Danckaert,
 J. Phys. B 41, (2008) 095402
\bibitem{Gelens2009b}
L.~Gelens, S.~Beri S, G.~Van~der~Sande, G.~Mezosi, M.~Sorel, J.~Danckaert, G.~Verschaffelt
Phys.~Rev.~Lett. 102 (2009) 193904
\bibitem{Beri08a}
S.~Beri, L.~Gelens, M.~Mestre, G.~Van~der~Sande, G.~Verschaffelt, A.~Scir\`e, G.~Mezosi, M.~Sorel, J.~Danckaert
Phys.~Rev.~Lett. 101 (2008) 093903
\bibitem{Gelens09a}
L.~Gelens, G.~Van~der~Sande, S.~Beri, J.~Danckaert
Phys.~Rev.~E 79 (2009) 016213
\bibitem{Melnikov91a}
V.~I.~Melnikov, Phys. Rep. 209 (1991) 1--71



\end{thebibliography}



\end{document}